\documentclass[10pt, conference, compsocconf]{IEEEtran}
\IEEEoverridecommandlockouts
\usepackage{cite}
\usepackage{graphicx}
\usepackage{soul}
\usepackage{color}
\usepackage{xcolor}
\graphicspath{ {./images/} }
\usepackage{amsmath,amssymb,amsfonts}
\usepackage{algorithmic}
\usepackage{graphicx}
\usepackage{textcomp}
\usepackage{xcolor}
\usepackage{hyperref}

\begin{document}

\title{Monitoring Efficiency of IoT Wireless Charging}

\author{\IEEEauthorblockN{Pengwei Yang, Amani Abusafia, Abdallah Lakhdari, and Athman Bouguettaya}
\IEEEauthorblockA{School of Computer Science\\
 The University of Sydney, Australia\\	
pyan8871@uni.sydney.edu.au,\\\{amani.abusafia,abdallah.lakhdari,athman.bouguettaya\}@sydney.edu.au}
}

\maketitle

\begin{abstract}
Crowdsourcing wireless energy is a novel and convenient solution to charge nearby IoT devices. Several applications have been proposed to enable peer-to-peer wireless energy charging.  However, none of them considered the \textit{energy efficiency} of the wireless transfer of  energy. In this paper, we propose an \textit{energy estimation framework} that predicts the actual received energy. Our framework uses two machine learning algorithms, namely XGBoost and Neural Network, to estimate the received energy. The result shows that the Neural Network model is better than XGBoost at predicting the received energy. We train and evaluate our models by collecting a real wireless energy dataset.

\end{abstract}

\vspace{-15pt}
\section{Introduction}


\textit{Internet of Things (IoT)} is a paradigm that enables everyday objects (i.e., \emph{things}) to connect to the internet and exchange data. IoT devices, such as smartphones and wearables, usually have augmented capabilities, including sensing, networking, and processing \cite{whitmore2015internet}. These advanced capabilities result in higher energy consumption and thereby require a longer battery life \cite{raptis2020wireless}. Increasing the battery capacity of IoT devices, on the other hand, faces several challenges, including safety, weight, cost, and recycling \cite{fang2018fair}. Additionally, traditional charging methods, such as charging cables or power banks, are inconvenient due to the high mobility of IoT users \cite{lakhdari2021fairness}. Therefore, charging  energy wirelessly  becomes an effective solution to improve battery endurance \cite{dhungana2020peer}\cite{abusafia2022quality}.\looseness=-1

The recent developments in wireless charging technologies  enable peer-to-peer wireless energy transfer between IoT devices \cite{YaoJessica2022WIES}. For example, Energous developed a technology to enable wireless charging up to a distance of 4.5 meters\footnote{www.energous.com}.
Recently, a framework was proposed to enable wireless energy-sharing services among IoT devices \cite{lakhdari2020Vision}\cite{abusafia2022services}. The framework leverages crowdsourcing energy from nearby IoT devices as energy services. \emph{Energy-as-a-Service (EaaS)} refers to the wireless transfer of energy from an IoT device (e.g. \textit{solar smartwatch}) to a nearby IoT device (e.g., \textit{smartphone})\cite{YaoJessica2022WIES}. An IoT device that offers energy is known as an \textit{energy provider}. Similarly, an IoT device that receives energy is known as an \textit{energy consumer}. Energy providers, such as smart textiles or solar watches, may \textit{harvest} energy from natural resources (e.g. body heat or physical activity) \cite{tran2019wiwear}\cite{sandhu2020towards}\cite{abusafia2022maximizing}. For instance, the PowerWalk kinetic energy harvester produces 10-12 watts of on-the-move power\footnote{www.bionic-power.com}. The harvested spare energy can be shared with nearby IoT devices as services.\looseness=-1

Recently, several mobile applications were developed in recent studies as first attempts to  enable peer-to-peer wireless energy services \cite{YaoJessica2022WIES}\cite{yang2022towards}. These applications allow smartphones to request energy services from nearby smartphones by size, e.g., 1000 mAh, or by time, e.g., to charge for the next 10 minutes. 
However, to the best of our knowledge, none of the existing apps considered the energy loss that occurs while transferring energy \cite{lakhdari2020Vision}\cite{abusafia2022services}. 
For instance, if a consumer request energy of 200 mAh, part of that energy will be lost in the transfer process depending on the used technology and the distance of the energy transfer \cite{dhungana2020peer}. Hence, they will receive an amount of energy less than what they requested. Therefore, we extend the work of \cite{YaoJessica2022WIES} and \cite{yang2022towards} to monitor the efficiency of the IoT wireless charging by building a wireless-transfer energy estimation framework. Our framework is used to \textit{estimate the actual amount of received energy} by a consumer. Our framework uses two machine learning models, i.e., XGBoost and Neural Network, to estimate the actual received energy. Both models consider two features in estimating the received energy: (1) the distance between the devices and (2) the duration of the energy transfer. We consider these features because different distances and durations impact the wireless transfer of energy\cite{yang2022towards}.\looseness=-1

\vspace{-5pt}
\section{Wireless Energy Estimation Framework}\label{sc}


\begin{figure}[!t]
    \centering
    \setlength{\abovecaptionskip}{-5pt}
     \setlength{\belowcaptionskip}{-70pt}
    \includegraphics[width=0.9\linewidth]{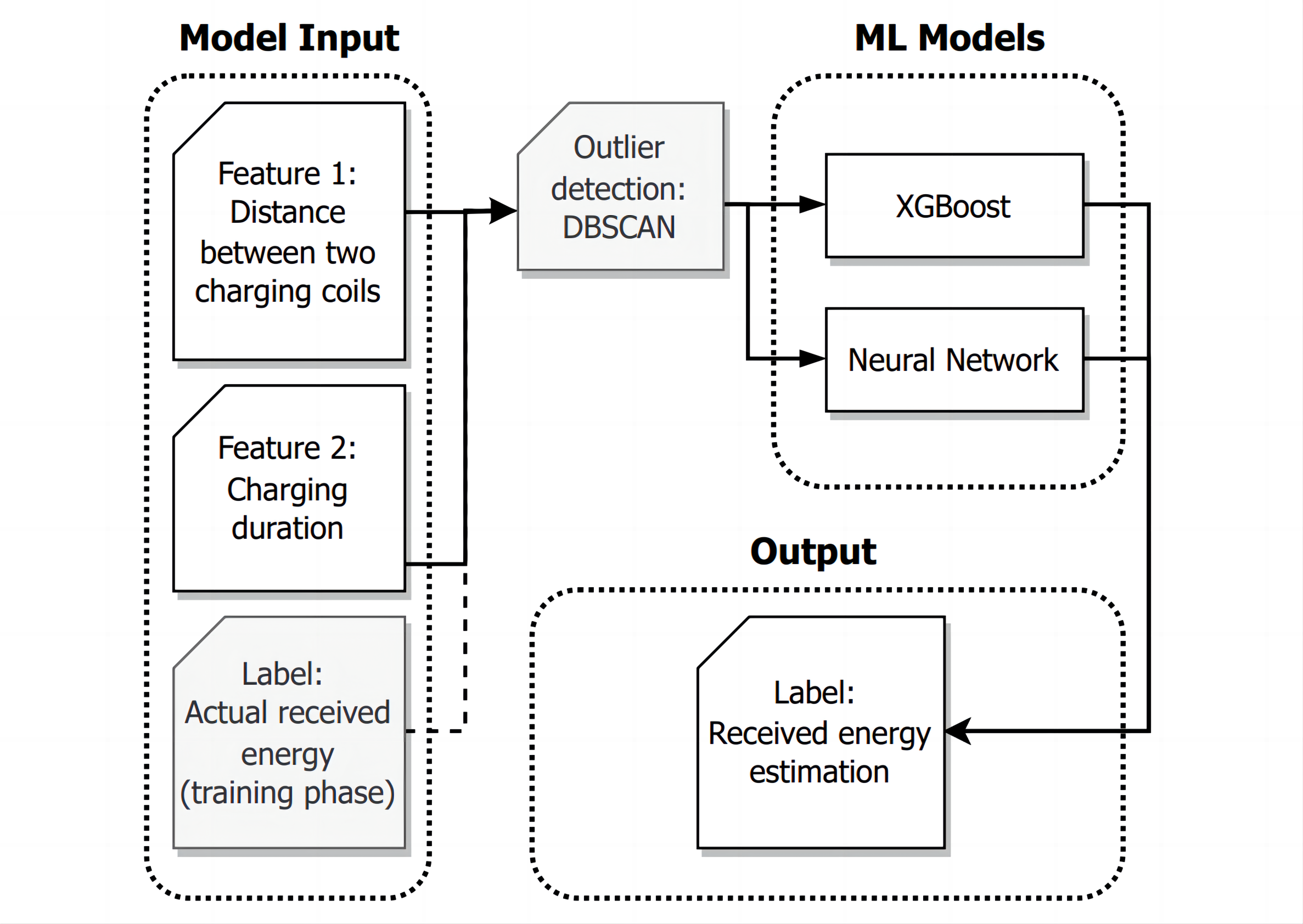}
    \caption{Wireless energy estimation framework}
    \vspace{-10pt}
    \label{framework}
\end{figure}
Our wireless energy estimation framework is presented in Fig.\ref{framework}. Specifically, our framework consists of four steps:
\vspace{-1pt}
\subsubsection{\textbf{Wireless Energy Transfer Dataset Collection}}

We collect a real wireless energy transfer dataset to train our machine-learning models. The data collection is done using  the app developed by \cite{yang2022towards}. The app allows consumers to connect to a provider via Bluetooth and then request energy based on size or time.  The app also monitors the energy-sharing process between a consumer and provider by recording their battery level every specific time interval ($mt$), e.g., every 5 seconds. The granularity of the monitoring time interval $mt$ is selected by the consumer.
A fine-grained time interval will give more detailed records of both the consumer and provider's battery levels. Moreover, a timestamp is used to synchronise the monitoring and recording of the energy transfer between the provider and consumer. Once an energy-sharing process is completed, the collected monitoring data will be pushed to the cloud.
The current version of the energy sharing application supports one-to-one energy transfer mode, i.e., a \textit{single} energy provider can deliver energy to only a \textit{single} energy consumer. 

\begin{figure}[!t]
    \centering
    \setlength{\abovecaptionskip}{-5pt}
    \includegraphics[width=0.7\linewidth]{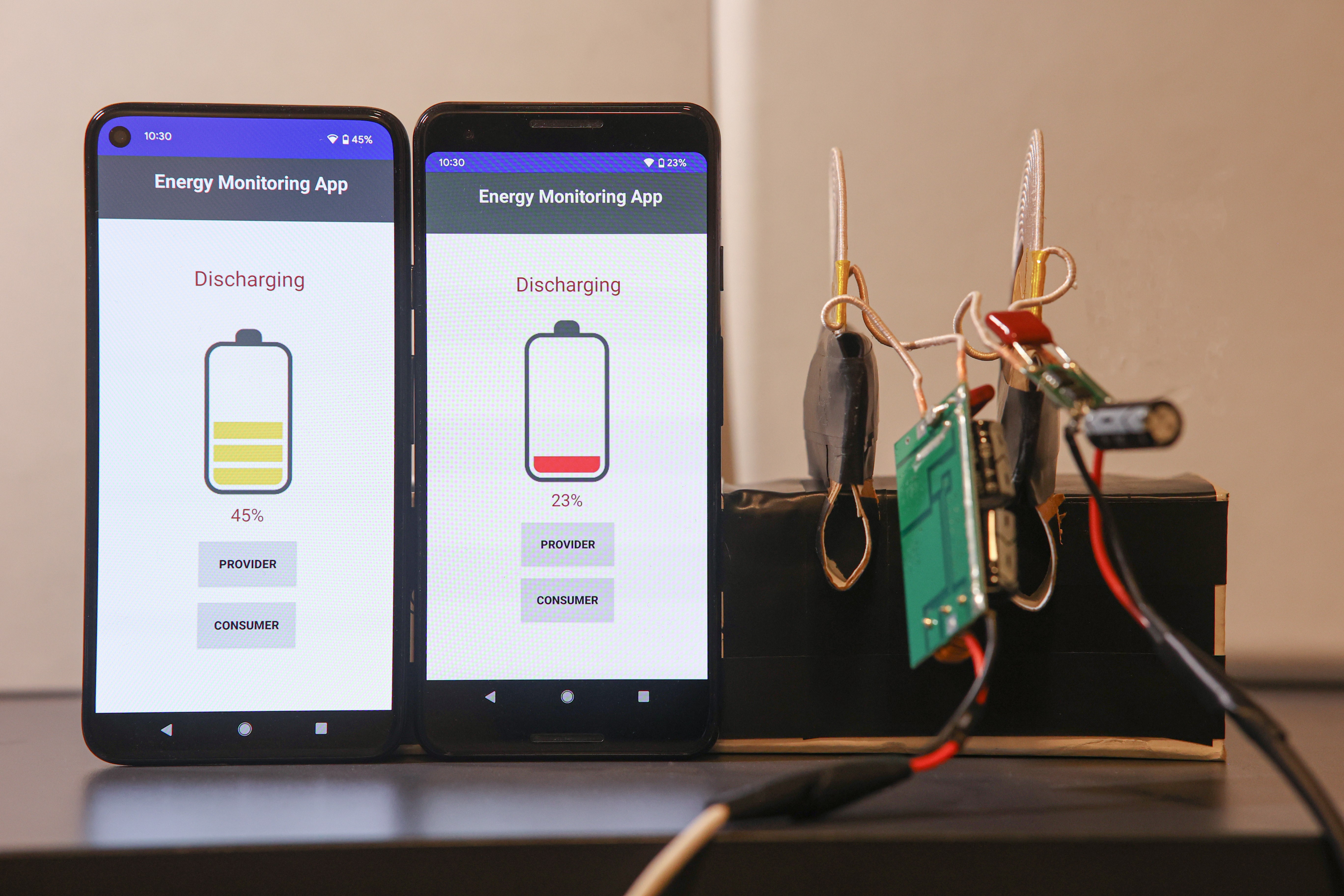}
    \caption{Hardware experiments setup for data collection
}
    \label{device}
    \vspace{-20pt}
\end{figure}

In our experiments, we used a Google Pixel 5 smartphone as the provider and a Google Pixel 3 as the consumer. We connected both the consumer and provider to  wireless charging coils (See Fig.\ref{device}). Moreover, in order to obtain accurate results, all experiments were conducted in the laboratory at temperatures around 20 degrees Celsius. Furthermore, we used the previously mentioned app to request energy by time. We set the charging period  to 30 minutes and the monitoring interval $mt$ to 1 minute. 
In order to conduct wireless charging at various distances between two coils, we created a sliding device (See Fig.\ref{device}). Users can use our device to share energy wirelessly over different distances. In our experiments,  
we collected datasets over different distances, i.e.,  1 cm, 1.5 cm, and 2 cm. We then repeated the experiments over each distance five times. Therefore, we have 450 data points (i.e., 5$\times$30$\times$3) in total. To sum up the collected dataset consist of five attributes: (1) distance between the coils, (2) charging duration, and (3) battery levels of the consumer and the provider every $mt$ time interval in mAh and percentage. In our experiments, $mt$ was 1 minute.\looseness=-1

\begin{figure*}[!ht]
    \centering
        \setlength{\abovecaptionskip}{-5pt}
    \includegraphics[width=0.8\linewidth]{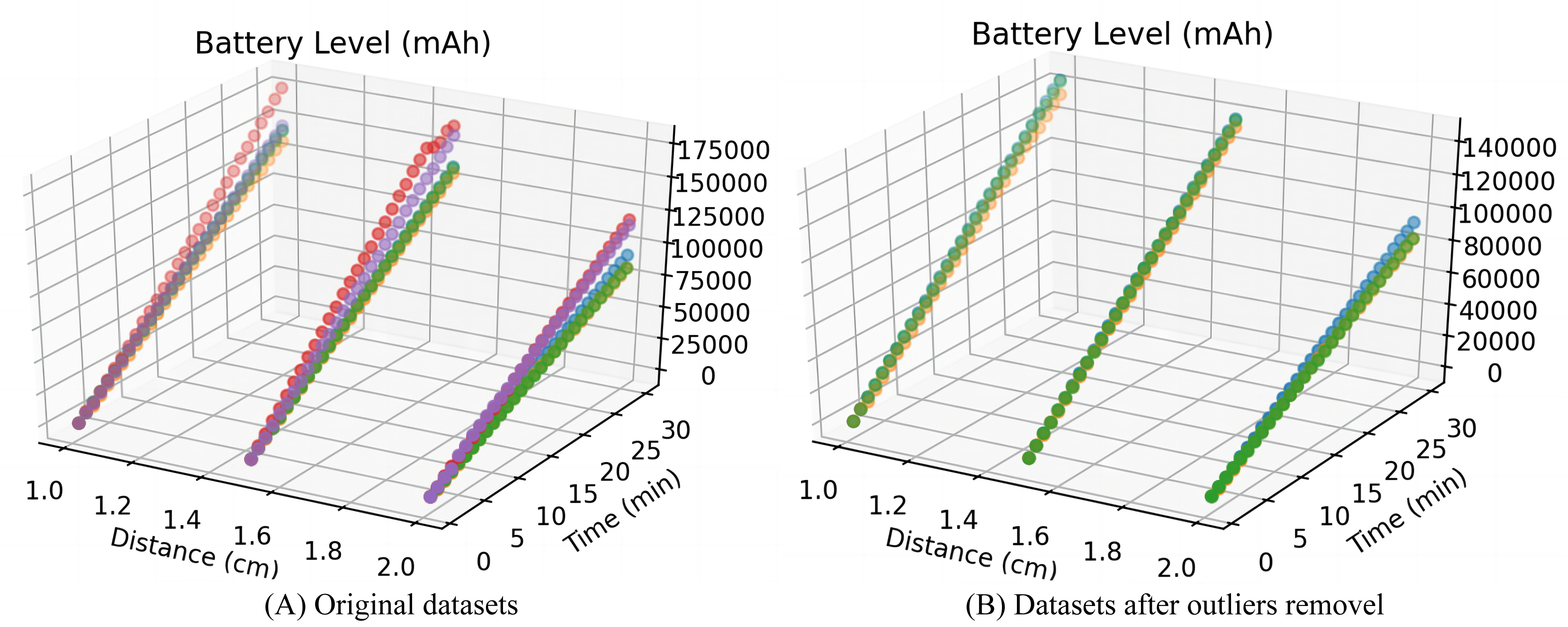}
    \caption{Training datasets visualization}
    \label{datasets}
        \vspace{-20pt}
\end{figure*}
\subsubsection{\textbf{Feature Selection}}

In our experiments, we consider the actual energy received by the consumer. Therefore,
We select the following attributes to train our models: (1) the charging duration (2) the consumer's battery level at each time, (3) and the distance between charging coils. The gathered dataset is visualized in Fig.\ref{datasets}(A). In training our models, we use the actual received energy as the label, while the other two variables as the selected features. As a result, we transform the objective of predicting the received energy  as a regression machine learning problem.\looseness=-1

\subsubsection{\textbf{Outlier Detection }}
The collected dataset had some abnormal charging battery levels that resulted in some of the data points being far away from the majority  of the data points (See Fig.\ref{datasets}(A)). As we used the Root Mean Square Error (RMSE) as our evaluation metric and due to RMSE sensitivity to outliers \cite{li2018predicting}, we use an outlier detection technique, namely, Density-Based Spatial Clustering of Applications with Noise (DBSCAN) \cite{ester1996density}, in order to detect outliers. In contrast to the statistical method that can detect anomalous points that are above and below a certain threshold (extremes), DBSCAN is able to discover data that occur infrequently (i.e., detect outlier points with a lower density) \cite{ccelik2011anomaly}. Moreover, there is a low likelihood of the occurrence of erroneous in the wireless charging process, which may result in abnormal values in the consumer battery level. Consequently, those anomalous data points have a lower density than normal data points. Therefore, we used DBSCAN to detect the outlier battery values. As aforementioned, a charging error may cause an anomalous rise in battery level. Hence, by focusing on the increase in the consumer battery level after the charging process, the problem is reduced to two dimensions: the distance between two charging coils and the increase in battery level. 
Fig.\ref{datasets}(B) depicts the training dataset excluding the outliers, which resulted in 270 data points in total. Furthermore, before feeding the data into our model, all of the target values, i.e., the battery level of the consumer, are divided by 1000 to reduce computational complexity. Hence, the unit of battery level is transformed from milliampere-hours (mAh) to ampere-hours (Ah) in our framework.\looseness=-1




\subsubsection{\textbf{Energy Estimation Models Training}}
We  use a conventional machine-learning method, namely, Extreme Gradient Boost (XGBoost), as our baseline \cite{chen2016xgboost}. The XGBoost algorithm is a boosting-based method for integrated learning. In our XGBoost model, we select the optimal hyperparameters using the GridSearch method in scikit-learn \cite{scikit-learn}.

Additionally, we use neural networks as our second model. Since the data distribution is not particularly complex, we construct a simple multi-layer perception with three hidden layers as our neural network model. As we have two input features, i.e., the distance between charging coils and the charging duration, we set the number of neurons in the input layer to two. Furthermore, as this is a regression machine learning problem, the number of neurons in the output layer is set to one. The result of our neural network model is a numerical value representing the estimated increase in the consumer battery level.

\vspace{-8pt}
\section{Evaluation}
 Since our energy estimation is a regression problem, the most straightforward way is to calculate the absolute gap between the predicted battery level increase and the actual battery level increase \cite{li2018predicting}. In this paper, we use a commonly used evaluation metric for regression problems, i.e., the Root Mean Square Error (RMSE) (See Eq.\ref{RMSE}). The numerator represents the summation of squares of the difference between the prediction and ground truth of each data point, while the denominator denotes the number of input data.

\begin{equation}\label{RMSE}
    RMSE = \sqrt{\frac{\sum(\hat{y_i}-y_i)^2}{n}}
\vspace{-5pt}
\end{equation}





\begin{figure}[!ht]
    \centering
    \setlength{\abovecaptionskip}{-5pt}
    \includegraphics[width=0.7\linewidth]{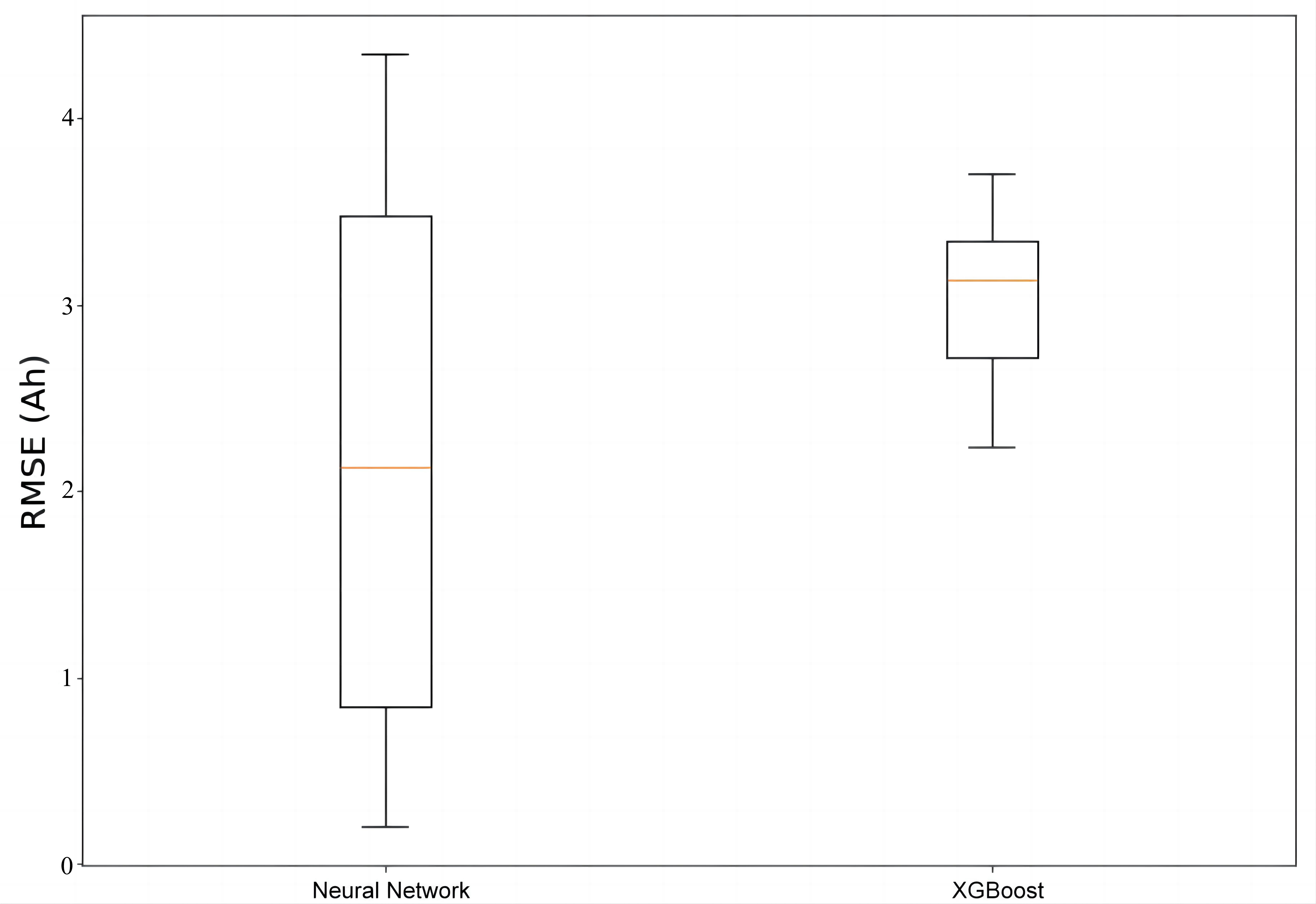}
    \caption{Box-plot of Root Mean Square Error (RMSE)
}
    \label{boxplot}
\end{figure}

The RMSE of XGBoost and neural network models are computed by ten-fold cross-validation. A comparison of both models' performances is represented in Fig.\ref{boxplot}.  The neural network model effectively reduces the RMSE score compared to the XGBoost model.
\vspace{-10pt}
    \vspace{-20pt}

\section{Demo SETUP}
At PERCOM 2023, we will demonstrate real cases of wireless energy sharing  between two smartphones over different distances. In this case, we use near-field wireless charging (i.e., inductive coupling wireless technique). The inductive coupling device consists of two coils, and the electrical energy is delivered based on the magnetic field (See Fig.\ref{device}). We will use a Google Pixel 5 smartphone as the provider and a Google Pixel 3 as the consumer. We connect both the consumer and provider to the wireless charging coils, and then we run a case similar to the previously mentioned scenario in Sec.\ref{sc}.
Additionally, interested visitors can also create their energy requests using our mobile application. They can also try different distances between the coils. Additionally, we will display a video of the entire process of using the devices and app to request, receive and monitor the energy-sharing process in real-time. The video is available at: youtu.be/8cYG6jscivE\looseness=-1



\section{Conclusion and Future Work}
In this paper, we presented our wireless energy estimation framework. The framework estimates the actually received energy considering the energy loss that occurs during the wireless energy transfer.  Our conducted experiments show that the neural network model outperforms  XGBoost model. In the future, we plan to study the impact of different coils' on energy transfer.\looseness=-1

\vspace{-5pt}
\section*{Acknowledgment}
This research was partly made possible by the LE220100078 grant from the Australian Research Council. The statements made herein are solely the responsibility of the authors.\looseness=-1

\bibliographystyle{IEEEtran}
\bibliography{main}
\end{document}